\documentclass[cite,aps,pra,amsmath, amsfonts, amssymb, superscriptaddress, twocolumn, showpacs]{revtex4}

\usepackage{graphicx}
\usepackage{braket}
\usepackage{wasysym}
\usepackage{setspace}
\usepackage{wasysym}
\usepackage{color}
\usepackage{amsmath}
\usepackage{amsfonts}
\usepackage{amssymb}

\def\a{s}
\def\b{s}
\renewcommand{\vec}[1]{\mathbf{#1}} 
\newcommand{\add}[1]{\if\a\b{{\color{red} #1}}\else{#1}\fi}
\newcommand{\comm}[1]{\if\a\b{{\color{blue} #1}}\else{#1}\fi}

\newcommand{\citeasnoun}[1]{Ref.~\onlinecite{#1}}

\newcommand{\chiany}[1]{\chi^{(#1)}}
\newcommand{\chitwo}{\chiany{2}}
\newcommand{\chithree}{\chiany{3}}

\renewcommand{\eqref}[1]{Eq.~(\ref{eq:#1})}

\newcommand{\eqrefrange}[2]{Eqs.~(\ref{eq:#1}--\ref{eq:#2})}

\newcommand{\secref}[1]{Sec.~\ref{sec:#1}}
\newcommand{\Secref}[1]{Section~\ref{sec:#1}}
\newcommand{\secreftwo}[2]{Secs.~\ref{sec:#1} and~\ref{sec:#2}}
\newcommand{\figref}[1]{Fig.~\ref{fig:#1}}

\newcommand{\Figrefrange}[2]{Figures~\ref{fig:#1}--\ref{fig:#2}}
\newcommand{\cav}{\textnormal{cav}}
\newcommand{\smax}{\textnormal{max}}
\newcommand{\crit}{\textnormal{crit}}

\begin{document}
\def\linefigwidth{0.5\textwidth}
\def\smalllinefigwidth{0.35\textwidth}
\def\smallerlinefigwidth{0.25\textwidth}
\def\largelinefigwidth{0.5\textwidth}

\title{Degenerate four-wave mixing in triply-resonant Kerr cavities}

\author{David Ramirez}
\affiliation{Department of Physics, Massachusetts Institute of Technology, Cambridge, MA 02139}
\author{Alejandro W. Rodriguez}
\affiliation{Department of Mathematics, Massachusetts Institute of Technology, Cambridge, MA 02139}
\affiliation{School of Science and Engineering, Harvard University, Cambridge, MA 02139}
\author{Hila Hashemi}
\affiliation{Department of Mathematics, Massachusetts Institute of Technology, Cambridge, MA 02139}
\author{J.~D.~Joannopoulos}
\affiliation{Department of Physics, Massachusetts Institute of Technology, Cambridge, MA 02139}
\author{Marin~Solja{\v{c}}i{\'{c}}}
\affiliation{Department of Physics, Massachusetts Institute of Technology, Cambridge, MA 02139}
\author{Steven~G.~Johnson}
\affiliation{Department of Mathematics, Massachusetts Institute of Technology, Cambridge, MA 02139}

\begin{abstract}
  We demonstrate theoretical conditions for highly-efficient
  degenerate four-wave mixing in triply-resonant nonlinear (Kerr)
  cavities. We employ a general and accurate temporal coupled-mode
  analysis in which the interaction of light in arbitrary
  microcavities is expressed in terms a set of coupling coefficients
  that we rigorously derive from the full Maxwell equations. Using the
  coupled-mode theory, we show that light consisting of an input
  signal of frequency $\omega_0-\Delta \omega$ can, in the presence of
  pump light at $\omega_0$, be converted with quantum-limited
  efficiency into an output shifted signal of frequency $\omega_0 +
  \Delta \omega$, and we derive expressions for the critical input
  powers at which this occurs. We find that critical powers in the
  order of 10mW assuming very conservative cavity parameters (modal
  volumes $\sim10$ cubic wavelengths and quality factors
  $\sim1000$. The standard Manley-Rowe efficiency limits are obtained
  from the solution of the classical coupled-mode equations, although
  we also derive them from simple photon-counting ``quantum''
  arguments. Finally, using a linear stability analysis, we
  demonstrate that maximal conversion efficiency can be retained even
  in the presence of self- and cross-phase modulation effects that
  generally act to disrupt the resonance condition.
\end{abstract}

\maketitle 


\section{Introduction}
\label{sec:intro}

For many years, researchers have used confinement of light for a long
time in a small volume (resonant cavities) to enhance light-matter
interactions such as optical nonlinearities, recently entering the
integrated-optics regime of smaller and smaller cavities with limited
sets of interacting modes. In such systems, careful design is required
to maximize the efficiency and minimize the power of a a given
nonlinear process such as frequency
conversion~\cite{Hashemi09,Rodriguez07:OE,Saleh91,Furst10,Ilchenko04,Lifshitz05,Freedman06,Bravo-AbadRo10,BurgessYi09,Bieler08,Hamam08,Morozov05,Kippenberg04,Bermel07,Bravo-Abad07,Fejer94,Hald01},
and the use of cavities can also lead to qualitatively new phenomena
such as bi/multistability
\cite{Felber76,Gibbs85,Cowan03,Hashemi09,Soljacic03:bisOL,Saleh91,Yanik03,Dutta98,Xu06,Soljacic02:bistable,Centeno00,Tanabe05,Notomi05,Parini07,Dorsel83,Billah90}. While
the use of cavities is known to enhance nonlinear effects, every
distinct nonlinear process requires a new analysis. In this paper, we
consider the problem of intra-cavity degenerate four-wave mixing
(DFWM): an electromagnetic cavity resonant at three frequencies
$\omega_0$ and $\omega_0 \pm \Delta\omega$, in which a third-order
($\chithree$) nonlinearity converts an input \emph{signal} at
$\omega_0 - \Delta\omega$ to an output \emph{shifted signal} at
$\omega_0 + \Delta\omega$ in the presence of an input \emph{pump} at
$\omega_0$.  The small-$\Delta\omega$ regime corresponds, for example,
to conversion between different channels in wavelength-division
multiplexing (WDM), similar to recent experimental studies of
nonlinear frequency conversion in cavities
\cite{Vahala03,Ferrera09,Ferrera08,Turner08,Heebner04,Absil00,Agha07,Broaddus09,Duchesne10};
it allows one to exploit structures like ring resonators
\cite{Saleh91} or photonic-crystal cavities
\cite{JoannopoulosJo08-book} that support closely spaced resonances,
and in \secref{qlc} we show that this regime supports stable
quantum-limited conversion at low signal powers for a critical pump
power.  Conversely, we show that the $\Delta\omega > \omega_0$ regime
generalizes our previous work on intra-cavity third-harmonic
generation (THG) \cite{Hashemi09,Rodriguez07:OE}, and in \secref{cc}
we show that this regime supports stable conversion with 100\%
efficiency at a critical pump and signal power. For example, with a
typical nonlinear material such as gallium arsenide (GaAs) and
reasonable cavity parameters (volume $\sim 10$ cubic wavelengths and
quality factors $\sim 1000$), we obtain critical powers in the
milliwatts (on the order of $10$mW) for both $\Delta\omega$ regimes.
The standard Manley--Rowe efficiency limits are considered from both a
simple photon-counting ``quantum'' argument \cite{Weiss57,Brown65} and
are also derived from purely classical coupled-mode equations
\cite{Haus84,Haus91} (\secref{qlc-cc}), where the latter also yield
stability information, critical powers, and other dynamics
(\secreftwo{qlc}{cc}). The coupling coefficients in these equations
are derived explicitly from the full Maxwell equations for arbitrary
microcavities (\secref{cmt}). We also show that the nonlinear dynamics
lead to additional phenomena, such as multistability and limit-cycle
(self-pulsing) solutions, similar to phenomena that were previously
shown for other nonlinear systems
\cite{Hashemi09,Drummond80,Grygiel92,Felber76}
(\secreftwo{qlc}{cc}). Finally, in \secref{alpha}, we consider the
effects of self- and cross-phase modulation (SPM and XPM), which
induce nonlinear shifts in the cavity frequencies: these must be
compensated by pre-shifting the resonances and also affect the
stability analysis (as we previously found for THG \cite{Hashemi09}).

\begin{figure}[t!]
\includegraphics[width=\columnwidth]{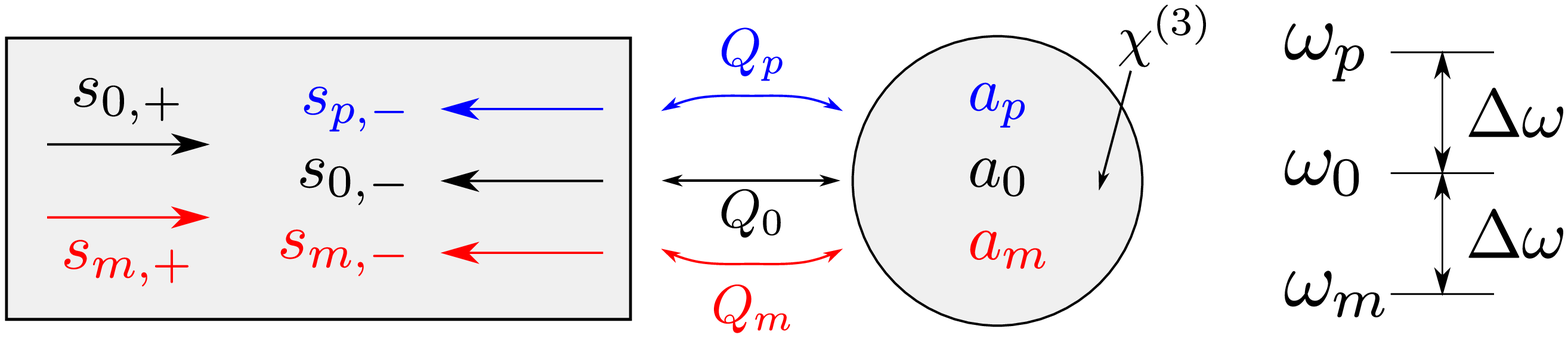}
\caption{(Color online) (\emph{Left}) Schematic for \emph{degenerate}
  four-wave mixing involving a coupled waveguide-cavity
  system. Dynamical variables for coupled-mode equations represent: a
  single input (output) channel (with incoming/outgoing field
  amplitudes $s_\pm$) coupled to a resonant cavity with three modes at
  frequencies $\omega_0$, $\omega_m = \omega_0 - \Delta \omega$ and
  $\omega_p = \omega_0 + \Delta \omega$ (and corresponding amplitudes
  $a_0$, $a_m$ and $a_p$).  The three resonant modes are nonlinearly
  coupled by a Kerr ($\chithree$) nonlinearity.  (\emph{Right})
  Diagram illustrating the relationship between the three resonant
  frequencies.}
\label{fig:fig1}
\end{figure}

Although nonlinear effects in electromagnetism are weak, it is well
known that confining light in a small volume and/or for a long time,
as in a waveguide or cavity, can both enhance the strength and modify
the nature of nonlinear phenomena \cite{Boyd92,Gibbs91}. Much previous
work in nonlinear frequency conversion has studied $\chitwo$ processes
(where there is a change in the susceptibility that is proportional to
the square of the electric field) such as second harmonic generation
(SHG)
\cite{Smith70,Drummond80,Wu87,Ou93,Paschotta94,Berger96,Zolotoverkh00,Maes05,Liscidini06,Schiller93:PhD,Ashkin66,Rodriguez07:OE,Furst10,Ilchenko04},
difference-frequency generation (DFG)
\cite{Burgess09:OE,Bieler08,Hamam08,McCutcheon10,Avetisyan99}, and
optical parametric amplification (OPA)
\cite{Baumgartner79,Yariv88,Moore95}. Studies of SHG in doubly
resonant $\chitwo$ cavities have demonstrated that 100\% conversion
efficiency is achieved at critical pump power, much lower than for SHG
in singly resonant cavities
\cite{Rodriguez07:OE,Paschotta94,Berger96,Zolotoverkh00,Maes05,Liscidini04,Dumeige06,Wu87,Ou93,Schiller93:PhD,Stivala10,Furst10,Ilchenko04}. Recent
studies of DFG in triply resonant $\chitwo$ cavities also showed the
existence of a critical relationship between pump and idler power that
results in optimal quantum-limited conversion \cite{Burgess09:OE},
with potential applications to terahertz
generation~\cite{BurgessYi09,Bravo-AbadRo10}. The existence of
quantum-limited frequency conversion can be predicted from the
Manley-Rowe relations, which govern the rates of energy transfer in
nonlinear systems \cite{Boyd92}. There has also been some recent work
on intra-cavity $\chithree$ third harmonic generation
\cite{Hashemi09}. (In a $\chithree$ medium, there is a change in the
refractive index proportional to the square of the electric field.) As
in SHG, THG in doubly resonant cavities has been shown to support
solutions with 100\% conversion efficiency, even when taking into
account nonlinear frequency-shifting due to SPM and XPM, as well as
interesting dynamical behavior such as multistability and limit cycles
(self-pulsing) \cite{Hashemi09}, with lower power requirements
compared to singly resonant cavities or nonresonant structures
\cite{Ashkin66,Smith70,Ferguson77,Brieger81,Bergquist82,Kozlovsky94,Dixon89,Collet91,Persaud90,Moore95,Schneider96,Mu01,Hald01,McConnell01,Dolgova02,Liu05,Scaccabarozzi06,Ferguson77}. Limit
cycles have been observed in a number of other nonlinear optical
systems, including doubly resonant $\chitwo$ cavities
\cite{Drummond80,Savage83}, bistable multimode Kerr cavities with
time-delayed nonlinearities \cite{Abraham82}, nonresonant distributed
feedback in Braggs gratings \cite{Parini07}, and a number of nonlinear
lasing devices \cite{Siegman86}.

In what follows, we extend the previous work on SHG, DFG and THG in
resonant cavities to the case of DFWM in $\chithree$ media. Four-wave
mixing is characterized by taking input light at frequencies
$\omega_1, \omega_2,$ and $\omega_3$ and producing light at frequency
$\omega_4 = \pm \omega_1 \pm \omega_2 \pm \omega_3$; degenerate
four-wave mixing, however, is restricted to the case where $\omega_1 =
\omega_3$ to generate $2\omega_1 - \omega_2$. Previous work has
studied FWM in the context of optical fibers
\cite{Levenson85,Shibata87,Inque92,Karisson98,Hansryd02} and even
matter waves \cite{Deng99}, as well as demonstrating the use of FWM in
applications such as phase conjugation
\cite{Feinberg82,Yariv88,Watanabe93:PTL,Agrawal02} and generation of
two-photon coherent states \cite{Yuen79,Levenson85,Slusher85}. While
there has been recent experimental work on intra-cavity FWM in
$\chithree$ media (degenerate or otherwise)
\cite{Vahala03,Ferrera09,Ferrera08,Turner08,Heebner04,Absil00,Agha07,Broaddus09,Duchesne10,PasPark10,Pasquazi10,Bartal06},
we are not aware of any detailed studies of the underlying theoretical
phenomena in general cavities. As we shall, see DFWM in
triply-resonant cavities shares many qualitative features with SHG,
DFG, and THG, including the existence of critical powers at which
optimal conversion efficiency is achieved as well as interesting
nonlinear phenomena such as limit cycles and multistability. As in
DFG, and unlike SHG or THG, there exist Manley-Rowe limitations on the
overall conversion efficiency. In \secref{qlc-cc}, we discuss the
corresponding relations governing four-wave mixing and illustrate
their implications for conversion efficiency. These relations can be
obtained classically through temporal coupled-mode
theory~\cite{Haus91,Haus84}, but they are more easily motivated and
understood from a quantum perspective \cite{Weiss57,Brown65}. Such
arguments have been employed before in the context of lasing
\cite{Huang03,Troccoli05}, RF circuits \cite{Maas03}, and other
nonlinear optics phenomena \cite{Boyd92}. In the case of intra-cavity
frequency conversion, we show how both perspectives yield limits on
conversion efficiency.

Several different approaches can be used to study nonlinear optical
systems.  Most directly, brute-force numerical simulations by a
variety of methods, such as finite-difference time-domain (FDTD)
\cite{Taflove00,JoannopoulosJo08-book}, offer the most general and
flexible technique, in that they can characterize phenomena involving
many degrees of freedom and going beyond the perturbative regime, but
such simulations are relatively slow and allow one to study only a
single geometry and excitation at a time. More abstract analyses are
possible in many problems because confinement to a waveguide or cavity
limits the degrees of freedom to the amplitudes of a small set of
normal modes, combined with the fact that optical nonlinearities are
typically weak (so that they can be treated as small perturbations to
the linear modes). For example, many nonlinear phenomena have been
studied in the context of co-propagating plane waves, in which the
amplitudes of the waves can be shown to satisfy a set of simple
ordinary differential equations (ODEs) in space (the slowly varying
envelope approximation) \cite{Boyd92}. More generally, however, it can
be shown that \emph{all} nonlinear problems coupling a finite set of
modes and satisfying certain fundamental principles such as
conservation of energy, regardless of the underlying wave equation
(e.g. electromagnetic or acoustic waves), can be described by a
\emph{universal} set of ODEs characterized by a small number of
coefficients, determined by the specific geometry and physics. This
approach, which has come to be known as temporal coupled-mode theory
(TCMT), dates back several decades \cite{Haus84,Suh04} and has been
applied to a large number of problems, from microwave transmission
systems \cite{Maas03} to the nonlinear intra-cavity problems (SHG,
DFG, and THG) mentioned above
\cite{Rodriguez07:OE,Burgess09:OE,Hashemi09,Suh04}. Likewise, we
employ TCMT in this paper to characterize the most general possible
behavior of intra-cavity DFWM systems, regardless of the nature of the
cavity. As reviewed elsewhere \cite{Haus84}, TCMT begins with the
purely linear system and breaks it into abstract components such as
input/output channels (e.g. waveguides or external losses) and
cavities, characterized by resonant frequencies and coupling rates
that depend on the geometry; it then turns out that the ODEs
describing such a system are completely determined by those parameters
once the constraints of conservation of energy, linearity,
time-invariance, and reciprocity (or time-reversal invariance) are
included, under the key assumption that coupling rates are slow
compared to the frequencies (i.e., strong confinement)
\cite{Haus84,JoannopoulosJo08-book}. Nonlinearities can then be
introduced as additional terms in these equations, without disturbing
the previously derived relationships, as long as the nonlinear
processes are also weak (i.e. nonlinear effects occur slowly compared
to the frequency), which is true in nonlinear optics
\cite{Boyd92}. Using these ODEs, the general possible behaviors can be
obtained (including the Manley-Rowe relations mentioned above), but to
obtain the specific characteristics of a particular geometry one then
needs a separate calculation to obtain the cavity
parameters. Properties of the linear modes such as frequencies and
lifetimes ($Q$) can be obtained by standard computational methods
\cite{Haus84,JoannopoulosJo08-book}. It turns out that the nonlinear
coefficients can also be obtained from the linear calculations, thanks
to the fact that the nonlinearities are weak: using perturbation
theory, expressions for the nonlinear coefficients as integrals of the
linear modes can be derived from Maxwell's equations. Such expressions
were previously derived for SHG and THG~\cite{Rodriguez07:OE}, and
also recently for DFG \cite{Burgess09:OE}.  Here, we derive both the
abstract TCMT equations and the specific nonlinear coupling
coefficients for DFWM in the Maxwell equations with $\chithree$
nonlinearities.

We begin apply the coupled-mode formalism to the case of DFWM in a
triply-resonant cavity in \secref{cmt}, to obtain the coupled-mode
equations of motion as well as explicit expressions for the nonlinear
coupling coefficients. We then briefly discuss general properties of
the conversion process in \secref{qlc-cc} and, using the standard
Manley-Rowe relations and simple photon-counting arguments, obtain
limits on the maximal efficiency of the system. In
\secreftwo{qlc}{cc}, we analyze the stability and dynamics of the
solutions to the coupled-mode equations obtained in \secref{cmt},
neglecting SPM and XPM effects, and demonstrate the existence of the
maximal conversion efficiencies obtained in \secref{qlc-cc}. Finally,
in \secref{alpha}, we briefly consider the effects of SPM and XPM
using a simple model to illustrate the qualitative behavior of the
system; in particular, we demonstrate the existence of stable, maximal
efficiency solutions even including SPM and XPM effects.

\section{Temporal coupled-mode theory}
\label{sec:cmt}

We consider the situation depicted schematically in \figref{fig1}: an
input/output channel coupled to a triply-resonant nonlinear
($\chithree$) cavity. Here, input light at $\omega_0$ and $\omega_m =
\omega_0 - \Delta \omega$ is converted to output light at a different
frequency $\omega_p = \omega_0 + \Delta \omega$, where $\Delta \omega$
determines the separation between the three frequencies. The
frequency-conversion process occurs inside the nonlinear cavity, which
supports resonant modes of frequencies $\omega_0$, $\omega_m$, and $
\omega_p$, and corresponding modal lifetimes $\tau_k$ (or quality
factors $Q_k = \omega_k \tau_k / 2$ \cite{Haus84}) describing the
overall decay rate ($1/\tau_k$) of the modes. In particular, the total
decay rate consists of decay into the output channel, with rate
$1/\tau_{s,k}$, as well as external losses (e.g. absorption) with rate
$1/\tau_{e,k}$, so that $1/\tau_k = 1/\tau_{s,k} + 1/\tau_{e,k}$. Note
that, to compensate for the effects of SPM/XPM, as described in
\cite{Hashemi09} and in \secref{alpha}, we will eventually use
slightly different cavity frequencies $\omega^{\cav}_k$ that have been
pre-shifted away from $\omega_k$.

It is most convenient to express the TCMT equations in terms of the
following degrees of freedom \cite{Haus84,JoannopoulosJo08-book}: we
let $a_k$ denote the time-dependent complex amplitude of the $k$th
mode, normalized so that $|a_k|^2$ is the electromagnetic energy
stored in this mode, and $s_{k,\pm}$ denote the time-dependent
amplitude of the incoming (+) and outgoing ($-$) wave, normalized so
that $|s_{k,\pm}|^2$ is the power in the $k$th mode. (In what follows,
we take $s_{p,+}=0$, corresponding to the up-conversion of light at
$\omega_0$ and $\omega_m$ to light at $\omega_p$, for $\Delta \omega >
0$. In order to study the alternative down-conversion process, one has
but to set $\Delta \omega <0$, in which case we effectively have
$\omega_p \to \omega_m$, as described below.)

The derivation of the linear TCMT equations, corresponding to
de-coupled modes $a_k$, has been given elsewhere \cite{Haus84}, and
the generalization to include nonlinearities has been laid out in
\citeasnoun{Rodriguez07:OE}. Here we introduce cubic nonlinearities
and make the rotating-wave approximation (only terms with frequencies
near $\omega_k$ are included in the equation of motion for $a_k$)
\cite{Rodriguez07:OE}. This yields the following general coupled-mode
equations:
\begin{align}
  \frac{da_0}{dt} = &\Big[ i \omega_0 (1-\alpha_{00}|a_0|^2 - \alpha_{0m}
  |a_m|^2 - \alpha_{0p} |a_p|^2) \nonumber \\ &- \left. \frac{1}{\tau_0} \right] a_0 - i
  \omega_0 \beta_0 a_0^* a_m a_p +
  \sqrt{\frac{2}{\tau_{s,0}}}s_{0,+} \label{eq:cme1} \\
\frac{da_m}{dt} = &\Big[ i \omega_m (1-\alpha_{m0}|a_0|^2 - \alpha_{mm}
|a_m|^2 - \alpha_{mp} |a_p|^2) \nonumber \\ &- \left. \frac{1}{\tau_m} \right] a_m -
i\omega_m \beta_m a_0^2 a_p^* +
\sqrt{\frac{2}{\tau_{s,m}}}s_{m,+} \label{eq:cme2} \\
\frac{da_p}{dt} = &\Big[ i \omega_p (1-\alpha_{p0}|a_0|^2 - \alpha_{pm}
|a_m|^2 - \alpha_{pp} |a_p|^2) \nonumber \\ &- \left. \frac{1}{\tau_p} \right] a_p  -
i\omega_p \beta_p a_0^2 a_m^* \label{eq:cme3} \\
s_{k,-} = &\sqrt{\frac{2}{\tau_{s,k}}} a_k - s_{k,+}. \label{eq:sminus}
\end{align}
As explained in \citeasnoun{Rodriguez07:OE}, the nonlinear
coefficients $\alpha_{ij}$ and $\beta_k$ depend on the specific
geometry and materials, and express the strength of the nonlinear
interactions. The $\alpha_{jk}$ terms describe self- and cross-phase
modulation effects which act to shift the cavity frequencies, while
the $\beta_k$ terms characterize the energy transfer (frequency
conversion) between the modes. As noted in
\citeasnoun{Rodriguez07:OE}, these terms are constrained by energy
conservation, which amounts to setting
$\frac{d}{dt}(|a_0|^2+|a_m|^2+|a_p|^2)=0$ (in the absence of external
losses), yielding the following relation:
\begin{equation}
  \omega_0 \beta_0^* = \omega_m \beta_m + \omega_p \beta_p.
\label{eq:betaconst}
\end{equation}
In the following sections, for simplicity, we neglect losses such as
linear absorption or radiation, i.e. we assume $\tau_{s,k} = \tau_k$,
and neglect nonlinear two-photon absorption, i.e. we assume
$\alpha_{ij}$ are strictly real (two-photon absorption effects can be
minimized by selecting materials less susceptible to such
processes). As noted below, these considerations do not qualitatively
change our results, but merely act to slightly decrease the overall
conversion efficiency once losses are included
\cite{Rodriguez07:OE,Hashemi09}.

The dependence of the coupling coefficients $\alpha_{ij}$ and
$\beta_k$ on the geometry of the system can be obtained via a simple
perturbative calculation involving the linear eigenmodes of the
cavity, as described in \citeasnoun{Rodriguez07:OE}. Carrying out this
procedure to first order in $\chithree$ yields the following coupling
coefficients:
\begin{widetext}
\begin{align}
  \beta_0 &= \frac{1}{8} \frac{\displaystyle\int{d^3\vec{x} \epsilon_0
      \chithree \left[(\vec{E}_0^* \cdot \vec{E}_0^*) (\vec{E}_m \cdot
        \vec{E}_p) + 2(\vec{E}_0^* \cdot \vec{E}_m)(\vec{E}_0^* \cdot
        \vec{E}_p)\right]}}{\displaystyle\left[\int{d^3\vec{x}
        \epsilon |\vec{E}_0|^2}\right] \left[\int{d^3\vec{x} \epsilon
        |\vec{E}_m|^2}\right]^{1/2} \left[\int{d^3\vec{x} \epsilon
        |\vec{E}_p|^2}\right]^{1/2}} \label{eq:beta0}\\ \beta_m &=
  \beta_p = \frac{1}{2}\beta_0^* \label{eq:betap} \\ \alpha_{jj} &=
  \frac{1}{8} \frac{\displaystyle\int{d^3\vec{x} \epsilon_0 \chithree
      \left[ |\vec{E}_j \cdot \vec{E}_j^* |^2 + |\vec{E}_j \cdot
        \vec{E}_j|^2\right]}}{\displaystyle\left[ \int{d^3 \vec{x}
        \epsilon |\vec{E}_j|^2}\right]^2} \label{eq:alphajj}\\
  \alpha_{jk} &= \frac{1}{8}
  \frac{\displaystyle\int{d^3\vec{x}\epsilon_0 \chithree \left[
        |\vec{E}_j \cdot \vec{E}_k|^2 + |\vec{E}_j \cdot
        \vec{E}_k^*|^2 +
        |\vec{E}_j|^2|\vec{E}_k|^2\right]}}{\displaystyle\left[\int{d^3\vec{x}\epsilon
        |\vec{E}_j|^2}\right]\left[\int{d^3\vec{x}\epsilon
        |\vec{E}_k|^2}\right]} \label{eq:alphajk}\\ \alpha_{kj} &=
  \alpha_{jk} \label{eq:alphakj},
\end{align}
\end{widetext}
where $\vec{E}_k$ is the electric field in the $k$th mode and the
denominators arise from the normalization of $|a_k|^2$.  As expected,
\eqrefrange{beta0}{betap} satisfy \eqref{betaconst}, where
\eqref{betaconst} was obtained by imposing energy conservation on the
TCMT equations without reference to the specific case of Maxwell's
equations.

There are six different $\alpha_{jk}$ parameters [three SPM
($\alpha_{jj}$) and three XPM ($\alpha_{jk}$) coefficients], and in
general they will all differ. However, from
\eqrefrange{alphajj}{alphajk} we see that they are all determined by
similar modal integrals, lead to frequency shifting of the cavity
frequencies, and all scale as $1/V$, where $V$ denotes a modal volume
of the fields \cite{JoannopoulosJo08-book}. Therefore, in the
following sections, we begin by neglecting the frequency-shifting
terms as in \citeasnoun{Hashemi09}, and then in \secref{alpha} we
study the essential effects of frequency shifting in the simplified
case where all the coefficients are equal ($\alpha_{jk} =\alpha$). Of
course, for a specific geometry one would calculate all coefficients
\eqrefrange{beta0}{alphakj}, but in this paper, we focus on the
fundamental physics and phenomena rather than the precise behavior of
a specific geometry.

\section{Quantum-limited vs. complete conversion}
\label{sec:qlc-cc}

\begin{figure}[t!]
\includegraphics[width=0.65\columnwidth]{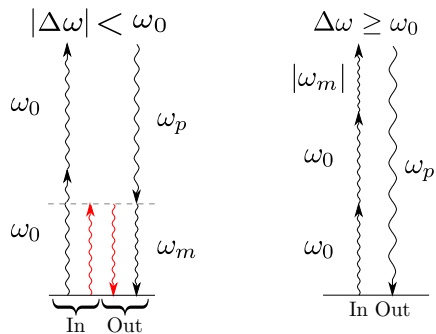}
\caption{(Color online) Diagram of nonlinear up-conversion process
  involving input light at $\omega_0$ and $\omega_m$ and output light
  at $\omega_p$ and $\omega_m$. The conversion efficiency of DFWM is
  determined by $\Delta \omega$, and photon energy conservation
  consideration (see text), leading to at least two different regimes
  of operation: (\emph{Left:}) for $|\Delta \omega | < \omega_0$, two
  $\omega_0$ pump photons and an signal $\omega_m$ photon are
  converted into two $\omega_m$ signal photons and an $\omega_p$
  photon. The input $\omega_m$ photon is only necessary to initiate
  the conversion process and emerges unchanged after the interaction
  (indicted by red).  (\emph{Right:}) for $\Delta \omega \geq
  \omega_0$, two incoming $\omega_0$ and a single $\omega_m$ photon
  are combined to produce an $\omega_p$ photon. In contrast to the
  previous regime, the $\omega_m$ photon is energetically needed to
  produce the $\omega_p$ photon.}
\label{fig:fig2}
\end{figure}

As described below, the DFWM process we consider here exhibits
drastically different behavior depending on the ratio of $\Delta
\omega$ to $\omega_0$. In particular, there exist at least two
distinct regimes of operation, corresponding to quantum-limited
($|\Delta \omega| < \omega_0$) and complete ($\Delta \omega \geq
\omega_0$) conversion. It turns out that, although our coupled-mode
formalism is entirely classical, the \emph{same} behaviors can be more
easily understood by considering photon interactions in a quantum
picture. Although this system is, of course, described by the general
Manley-Rowe relations, which can be derived from both
classical~\cite{Haus84,Haus91,Boyd92} and
quantum~\cite{Weiss57,Brown65} arguments similar to those here, it is
useful to review a basic picture of such limits and their physical
consequences for the specific case of intra-cavity DFWM.

Our focus in this manuscript is the up-conversion process (or
interaction) corresponding to taking input light at frequencies
$\omega_0$ and $\omega_m$ and generating output light at frequency
$\omega_p$.  Therefore, an appropriate figure of merit is the ratio of
the output power in the $\omega_p$ mode to the total input power,
which we define as the absolute efficiency $\eta =
|s_{p,-}|^2/(|s_{0,+}|^2+|s_{m,+}|^2)$.

As described in the previous section, the coupled-mode equations
[\eqrefrange{cme1}{sminus}] follow from very general and purely
classical considerations. The same considerations yield relationships
between the frequencies and coupling coefficients of the problem, such
as frequency conservation ($\omega_m + \omega_p = 2 \omega_0$) and
energy conservation ($\omega_m \beta_m + \omega_p \beta_p = \omega_0
\beta_0^*$). Additional conservation rules that are perhaps best
understood from quantum arguments, such as photon energy ($\hbar
|\omega|$) conservation and standard $\chithree$ selection rules
\cite{Boyd92}, also play a substantial role in the physics of
nonlinear frequency conversion.  In the case of the DFWM up-conversion
process considered here, $\chithree$ selection rules imply that
nonlinear interactions can only be initiated if there exist at least
three input photons: $2\omega_0$ photons and one $\omega_m$ photon.

In the $|\Delta \omega| < \omega_0$ regime, there are at least two
important features that can be understood from the above relations:
First, depletion of the signal input power ($s_{m,+}$) is impossible,
leading to a conversion efficiency $\eta < 1$. Second, in order to
maximize the total conversion efficiency, one desires $s_{m,+}$ to be
as small as possible. These features can be understood by considering
a simple picture of the nonlinear photon--photon interaction, as
follows. From the DFWM $\chithree$ selection rule \cite{Boyd92}, it
follows that the creation of an $\omega_p$ photon is accompanied by
the destruction of two $\omega_0$ photons and one $\omega_m$
photon. The latter, along with photon energy conservation, leads to
the process considered in \figref{fig2} (left), in which two
$\omega_0$ photons and an $\omega_m$ photon interact to yield two
$\omega_m$ photons and an $\omega_p$ photon. From the figure, and
since $2\omega_0 > \omega_p$, one can see that the incident $\omega_m$
photon (depicted in red) is merely required by the $\chithree$
selection rule to initiate the interaction, and emerges unmodified,
accompanied by an $\omega_p$ photon and an additional $\omega_m$
photon. Thus, it is clear that the input $\omega_m$ photon does not
actively participate in the energy transfer and therefore merely
reduces the maximum possible conversion efficiency. This implies that
one desires a minimal input signal power to initiate the
up-conversion. Effectively, the incident $\omega_m$ photons are
amplified by the conversion process (a similar amplification effect is
a crucial component in other nonlinear interactions, such as OPAs in
$\chitwo$ media \cite{Armstrong62,Boyd92,Choi97}). In addition, it is
clear that complete depletion of the signal photons, i.e. $s_{m,-} =
0$, is not possible for non-zero $s_{m,+}$, and therefore the
conversion efficiency must be less than 100\% (since the total input
power is conserved). No such restriction is placed on $s_{0,-}$, and
therefore we expect that maximal efficiency will be obtained for
arbitrarily low signal power and complete depletion of the pump power,
i.e. $s_{0,-} = 0$.

Based on these arguments, we can predict the maximal efficiency of the
conversion process by considering the ratio of the energy of the
output $\omega_p$ photon ($\hbar \omega_p$) to the energy of the three
input photons [$\hbar (2\omega_0 + \omega_m)$]. Since the $\omega_m$
photons can be provided with arbitrarily low amplitude, we therefore
expect maximal efficiency to be achieved upon neglecting their
contribution, i.e. we predict a maximal efficiency of:
\begin{equation}
  \label{eq:etamaxpr}
  \eta_{\smax}(|\Delta \omega< \omega_0|) = \frac{\hbar \omega_p}{2 \hbar \omega_0} = \frac{\omega_p}{2\omega_0}.
\end{equation}
Note that this efficiency depends only on the ratio of $\Delta \omega$
to $\omega_0$ and $\hbar$ cancels, so it should appear in the
classical limit as well. As we shall see in \secref{qlc}, this
prediction is verified analytically by examining the steady-state
solution of our coupled-mode equations.

In the $\Delta \omega \geq \omega_0$ regime, the conversion process is
fundamentally different and, in particular, complete depletion of the
$\omega_m$ and $\omega_0$ photons is possible, leading to 100\%
conversion efficiency. Basically, because $\omega_p > 2\omega_0$ in
this case, no additional photons are required to satisfy photon energy
conservation, yielding the nonlinear interaction process depicted in
\figref{fig2} (right), where two input $\omega_0$ photons and an
$\omega_m$ photon combine to produce an $\omega_p$ photon. Note that
now the input $\omega_m$ photon actively participates in the energy
transfer, in contrast to the $|\Delta \omega| < \omega_0$ regime,
leading to a maximal conversion efficiency occurring when $s_{0,+}$
and $s_{m,+}$ are both non-zero. Furthermore, since $\omega_p$ is now
the only product of the interaction, we expect that complete depletion
of both the pump and signal powers, $s_{0,-} = s_{m,-} = 0$, should be
possible, leading to 100\% conversion efficiency. As before, this can
also be quantified by comparing the ratio of the output energy ($\hbar
\omega_p$) to the input energy [$\hbar (2 \omega_0 + |\omega_m|)$]
(note that now the energy of $\omega_m$ photon is $\hbar |\omega_m|$),
and the result follows from the fact that $2\omega_0 + |\omega_m| =
\omega_p$. Again, we shall see in \secref{cc}, this prediction is
validated analytically and directly from the coupled-mode equations,
yielding also the critical input powers at which 100\% conversion is
achieved.

In this section, we made a number of predictions based on very general
arguments relying on a quantum interpretation of the nonlinear
interactions, allowing us to obtain predictions of maximal conversion
efficiency. Our final results, of course, contained no factors of
$\hbar$ and it is therefore not surprising that we recover the same
results (albeit with more detail, e.g. predictions of the values of
critical powers) in the ensuing analysis of the purely-classical
coupled-mode equations. Nevertheless, the heuristic quantum picture of
\figref{fig2} has the virtue of being simple and revealing, while the
classical derivation is more complicated (although more
quantitative). Similar quantum arguments have also proven useful in
other contexts, such as in many problems involving classical radiation
\cite{Jackson98}, or the recently-studied problem of optical
bonding/anti-bonding in waveguide structures \cite{Povinelli05}.

\section{Coupled-Mode Analysis}
\label{sec:cma}
In order to gain a simple understanding of the system, we shall first
consider frequency conversion in the absence of self- and cross-phase
modulation, i.e. $\alpha_{jk} = 0$. The nonzero-$\alpha$ case will be
considered in \secref{alpha}. \Secref{qlc} focuses on the $|\Delta
\omega| < \omega_0$ regime, whereas \secref{cc} focuses on the
$\Delta\omega \geq \omega_0$ regime. In both cases, we describe the
solutions to the coupled-mode equations [\eqrefrange{cme1}{cme3})] in
the steady state, including the statbility of these solutions and
their dependence on the cavity parameters.

\subsection{$|\Delta \omega| < \omega_0$ regime: Limited conversion}
\label{sec:qlc}

Although the analysis in this section is general, for the purposes of
plotting results we choose the specific parameters: $\alpha_{jk}=0$,
$\tau_0=\tau_m=\tau_p=100/\omega_0$, $\beta=10^{-4}$, and $\Delta
\omega = 0.05 \omega_0$. The qualitative results remain unchanged as
these parameters are varied, provided that the $Q$ are large enough
such that mode overlap is minimal as required by CMT. The influence of
varying these parameters is discussed further at the end of the
section.

To understand the stability and dynamics of the nonlinear coupled-mode
equations in the quantum-limited regime, we apply the standard
technique of identifying the fixed points of \eqrefrange{cme1}{cme3}
and analyzing the stability of the linearized equations around each
fixed point \cite{Tabor89}. A fixed point is given by a steady-state
solution where the mode amplitudes vary as $a_k(t) = A_k e^{i \omega_k
  t}$, with the $A_k$ being unknown constants. Plugging this
steady-state ansatz into \eqrefrange{cme1}{cme3}, we obtain three
coupled polynomial equations in the parameters $A_0, A_m, A_p,
s_{0,+},$ and $s_{m,+}$. These polynomials were solved using
Mathematica to obtain the mode energies $|A_k|^2$, which are then used
to calculate the efficiency $\eta = |s_{p,-}|^2 / (|s_{0,+}|^2 +
|s_{m,+}|^2)$. The phases of the $A_k$ can be easily determined from
the steady-state equations of motion; $A_0$ and $A_m$ acquire the
phases of $s_{0,+}$ and $s_{m,+}$ respectively, while the phase of
$A_p$ is that of $\beta_p A_0^2 A_m^*$ rotated by $\pi/2$. Without
loss of generality, $s_{0,+}$ and $s_{m,+}$ can be chosen to be real.

\begin{figure}[t]
\includegraphics[width=\columnwidth]{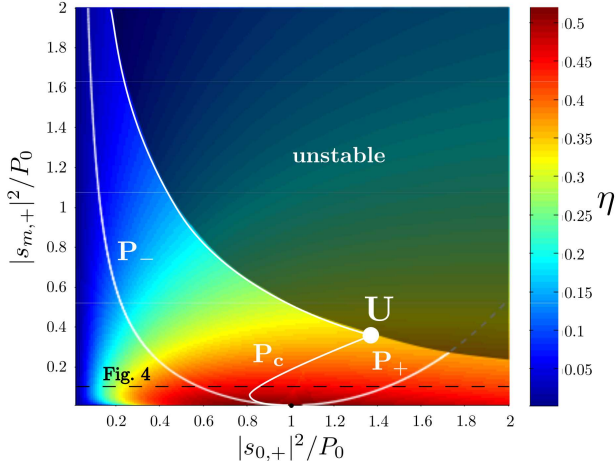}
\caption{(Color online) Color plot of the steady-state conversion
  efficiency $\eta = |s_{p,-}|^2 / (|s_{0,+}|^2 + |s_{m,+}|^2)$ as a
  function of input power $|s_{0,+}|^2$ and $|s_{m,+}|^2$, for a
  system consisting of $\Delta \omega = 0.05 \omega_0$, $\beta =
  10^{-4}$, and $\tau_0 = \tau_p = \tau_m = 100$. Both powers are
  normalized by the critical power $P_c(|s_{m,+}|^2\to 0) = P_0 =
  2/\tau_0 |\beta| \sqrt{\tau_m \tau_p |\omega_m \omega_p|}$ (black
  dot). The shaded region indicates that the solution is unstable. The
  curves $P_{\pm}$ indicate the powers at which depletion of the
  $\omega_0$ input light is achieved, i.e. $s_{0,-} = 0$; the critical
  power $P_c(|s_{m,+}|^2)$ is defined as the total input power that
  yields the highest \emph{stable} efficiency for any given
  $|s_{m,+}|^2$. The dash line is the cross-section shown in
  \figref{fig4}.}
\label{fig:fig3}
\end{figure}

In general, this system has either one or three solutions, only one of
which is ever stable. The stability and efficiency of this solution
are shown in \figref{fig3} for the specific parameters mentioned
above. We observe that maximal conversion efficiency is obtained in
the limit as input signal power $s_{m,+}$ is reduced to zero,
consistent with the discussion in the previous section. To obtain the
maximum efficiency and the corresponding critical input powers,
complete depletion of the pump ($\omega_0$) photon is required,
i.e. $s_{0,-} = 0$ (note that one cannot require depletion of the
signal photon, for the reasons discussed in the previous section). We
find that the maximum efficiency $\eta_\smax$ is obtained at
$|s_{0,+}^\crit|^2 = P_0$ as $|s_{m,+}|^2 \to 0$, where:
\begin{align}
  P_0 &= \frac{4}{\tau_0 |\beta_0| \sqrt{\tau_m \tau_p |\omega_m
      \omega_p|}} \label{eq:p0} \\ \eta_\smax &= \frac{\omega_p}{2\omega_0}
  = \frac{1}{2}\left(1+ \frac{\Delta
      \omega}{\omega_0}\right) \label{eq:etamax},
\end{align}
Note that \eqref{etamax} is identical to the value predicted in the
previous section. In the important case of narrow-band conversion,
$|\Delta \omega | \ll \omega_0$, the maximum efficiency is
approximately $50\%$. (however, this is relative to the pump power---
compared to the input signal alone, the output signal is amplified to
an arbitrary degree). If $Q_0 \sim Q_m \sim Q_p$, then, as in THG
\cite{Hashemi09}, the critical power scales as $V/Q^2$, where $V$ is
the modal volume (recall that $\beta \sim 1/V$).

As $\Delta \omega \to \omega_0$, the maximum efficiency approaches
unity, i.e. 100\% conversion can be achieved in the limit. This limit
is reminiscent of second-harmonic generation, since $\omega_p = 2
\omega_0$. However, the interaction process is fundamentally different
from the standard ($\chitwo$) SHG in a number of ways. First, one is
converting DC ($\omega_m \approx 0$) light and $\omega_0$ pump light
into $2\omega_0$. Second, the stability of this solution (described
below) is quite different from that of SHG
\cite{Drummond80,Savage83,Grygiel92}. Finally, the critical power in
this case, $P_0$, diverges as $1/\sqrt{1 - (\Delta \omega /
  \omega_0)^2}$ for $\Delta \omega$ near $\omega_0$. However, $\Delta
\omega$ close but not equal to $\omega_0$ yields a reasonable $P_0$:
for example, $\Delta \omega = 0.95\omega_0$ yields efficiency
$\eta=0.975$ with a critical power roughly three times the critical
power for $\Delta \omega $ near zero. Because this near-``SHG''
situation involves coupling resonances at very different frequency
scales, it is reminiscent of using $\chitwo$ DFG to produce THz from
infrared \cite{Burgess09:OE}.

\eqrefrange{p0}{etamax} are only valid in the limit $|s_{m,+}|^2\to
0$, which is ideal from an efficiency perspective. However, it is
interesting to consider the system for non-infinitesimal $s_{m,+}$, in
which case we solve for the input power that yields a stable solution
with maximal efficiency for a given $s_{m,+}$. We denote this input
power by $P_c(|s_{m,+}|^2) = |s_{0,+}^\crit|^2 + |s_{m,+}|^2$, where
$|s_{0,+}^\crit|^2$ (a function of $|s_{m,+}|^2$) is defined to be the
pump power required to achieve maximum, stable conversion efficiency
for a given signal power $|s_{m,+}|^2$. As seen in \figref{fig3}, this
efficiency is always $\leq \eta_\smax$, and $P_c \to P_0$ as $s_{m,+}
\to 0$. In the non-zero $|s_{m,+}|^2$ regime, $P_c$ does not
correspond to complete depletion of the pump. Requiring pump depletion
($s_{0,-} = 0$) for a given signal power $|s_{m,+}|^2$ yields two pump
powers, which we label $P_{\pm}(|s_{m,+}|^2)$. $P_+(|s_{m,+}|^2)$ does
indeed provide a solution with maximal efficiency, however this
solution is always unstable. As seen from \figref{fig3}, only for
small signal power $s_{m,+}$ does depletion of the pump lead to
maximal efficiency.

\begin{figure}[t!]
\includegraphics[width=\columnwidth]{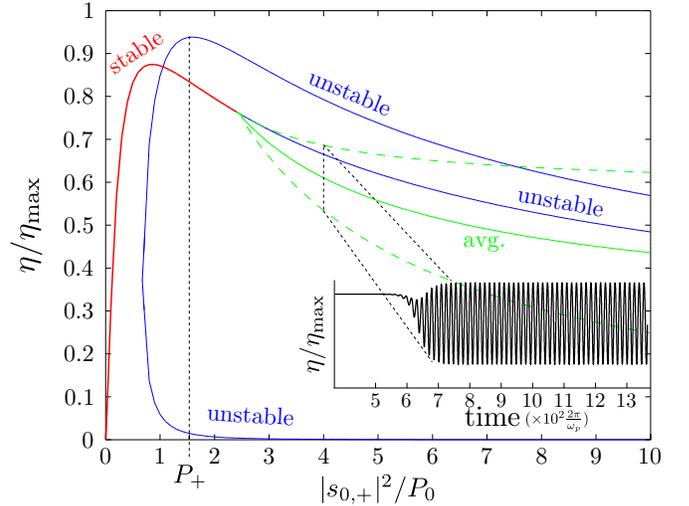}
\caption{(Color online) Bifurcation diagram of the steady-state
  efficiency $\eta$, normalized by the quantum-limited maximum
  efficiency $\eta_\smax= \frac{1}{2}\omega_p/\omega_0$, as a function
  of $|s_{0,+}|^2$, normalized by $P_0$, for signal power $|s_{m,+}|^2
  = 0.1 P_0$ (indicated by the black dashed line of
  \figref{fig3}). Red/blue correspond to a stable/unstable solution
  (note that the two bifurcating solutions are always unstable). The
  green dashed line illustrates the bounds of the limit cycles
  obtained from time domain simulations, where the solid green line
  yields the average over the cycle. (\emph{Inset:}) Efficiency as a
  function of time in units of the period $T_p = 2\pi/\omega_p$ in a
  regime where there exists a limit cycle.}
\label{fig:fig4}
\end{figure} 

\begin{figure}
\includegraphics[width=\columnwidth]{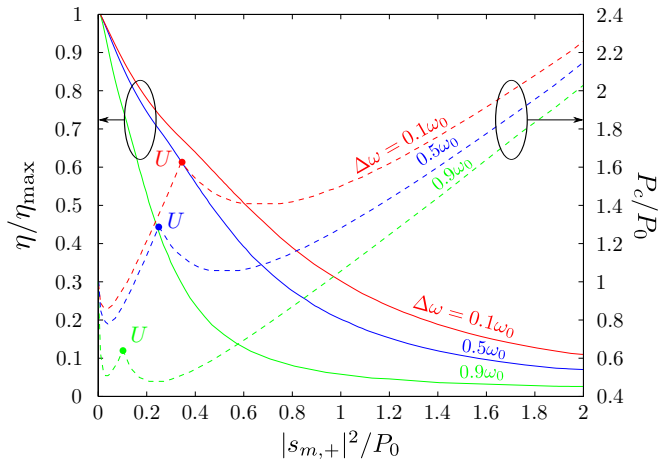}
\caption{(Color online) Plot of the steady-state efficiency $\eta$
  (solid lines) along the critical solution [total input power
  $P_c(|s_{m,+}|^2) = |s_{m,+}|^2 + |s^c_{0,+}|^2$ that yields the
  maximum efficiency for a given $|s_{m,+}|^2$ (solid white curve of
  \figref{fig3})] and the value of $P_c$ (dashed lines) as a function
  of $|s_{m,+}|^2$, normalized by $P_0$, for three different values of
  $\Delta \omega$: $0.1 \omega_0$ (red), $0.5 \omega_0$ (blue), and
  $0.9 \omega_0$ (green). The kinks in the $P_c$ curves correspond to
  the point $U$ where $P_c$ reaches the region of instability (see
  \figref{fig3}). The coupling lifetimes $\tau$ and coefficient
  $\beta$ of the system are equivalent to those of \figref{fig3}.}
\label{fig:fig5}
\end{figure}

In general, to obtain the largest efficiency while retaining
stability, one would aim to operate with low signal power
$|s_{m,+}|^2$ and use a pump power near the critical power $P_0$ given
in \eqref{p0}. However, it is interesting to consider the unstable
solutions, because they turn out to be related to limit cycles. As
mentioned above, the system contains either one or three steady-state
solutions for given input powers. \figref{fig4} plots these stable and
unstable solutions as a function of pump power $|s_{0,+}|^2$ at fixed
signal power $|s_{m,+}|^2 = 0.1 P_0$, corresponding to the horizontal
dashed line in \figref{fig3}. For low input pump power $|s_{0,+}|^2$,
the system has a single steady-state solution; as the pump power is
increased, the system experiences a bifurcation yielding two unstable
solutions. As mentioned above, the higher efficiency solution emerging
from the bifurcation achieves a maximum corresponding at $|s_{0,+}|^2
= P_+$, coinciding with complete depletion of the pump ($s_{0,-} =
0$), but this maximal efficiency solution is always unstable; note
that there may be a stable solution at $|s_{0,+}|^2 = P_+$, but the
stable solution will have a lower efficiency than the maximal,
unstable solution, as shown in \figref{fig4}. Furthermore, the
original stable solution eventually becomes unstable as the pump power
is increased (this can occur before or after the bifurcation,
depending on the system parameters); this onset of instability
coincides with the onset of limit cycles, stable
oscillating-efficiency solutions. An example of these limit cycles are
shown in \figref{fig4}, where the green dashed lines indicate the
bounds of the oscillations and the solid green line gives the
average. The limit cycles are plotted as a function of time in the
inset of \figref{fig4}. The limit cycles shown here were obtained by
numerically time-evolving the coupled-mode equations. In general, we
find that these limit cycles oscillate with a period proportional to
$\tau_p$.

\Figrefrange{fig3}{fig4} describe a system corresponding to a
particular set of values for the parameters $\Delta \omega$ and
$\tau_k$. Qualitatively, the most important features of the figures
remain largely unchanged as these parameters are varied. Basically,
there exist at most three solutions to the coupled-mode equations, one
of which has a finite region of stability as a function of $s_{0,+}$
and $s_{m,+}$, with the general shape that is shown in \figref{fig3},
and two others that are always unstable and bifurcate at a finite
$s_{0,+}$. There are however, some differences to note: First, as
$\Delta \omega$ increases from 0, the maximum steady-state efficiency
also increases, asymptoting to $\eta = 1$ as $\Delta \omega \to
\omega_0$. This was obtained analytically and is quantified in
\eqref{etamax}. Unfortunately, we find that as $\Delta \omega$
increases, the region of instability in \figref{fig3} also increases,
and furthermore, the conversion efficiency at finite $s_{m,+}$ also
drops off more rapidly. (In particular, we observe in the ``SHG''
limit of $\Delta \omega \to \omega_0$, the system becomes largely
unstable except for very low signal powers.) These tendencies are
depicted in \figref{fig5}, which plots $P_c(|s_{m,+}|^2)$ and the
corresponding conversion efficiency for different values of $\Delta
\omega$. The kinks observed in the plots of $P_c$ are due to the
discontinuity in the slope of the $P_c$ curve as it reaches the region
of instability, corresponding to the point $U$ in \figref{fig3}.

Varying $\tau_k$ does not affect the maximum possible efficiency and
also leaves \figref{fig3} qualitatively unchanged, changing only the
scale of the critical input power $P_0$. The stability of the system
however, does depend on the relative lifetimes of the cavity modes. In
particular, the stability depends largely on the ratio $\tau_0 /
\tau_p$, and decreases weakly as $\tau_m$ increases with respect to
either $\tau_0$ or $\tau_p$. This makes sense since, as argued in
\secref{qlc-cc}, the $\omega_m$ photons do not actively participate in
the energy transfer. (A similar dependence on the ratio of the
lifetimes was also observed in the case of THG \cite{Hashemi09}.) More
quantitatively, we follow the position of the point $U$ (the point
where $P_c$ reaches the region of instability) as the $\tau_k$ are
varied. Assuming equal modal lifetimes ($\tau_0 = \tau_m = \tau_p$ as
in \figref{fig3}), we find that $U$ lies at critical input powers
$|s_{0,+}|^2 \approx 1.28 P_0$ and $|s_{m,+}|^2 \approx 0.35
P_0$. Increasing $\tau_0/\tau_p$, from $1$ to $10$, we find that $U$
moves to $|s_{0,+}|^2 \approx 10 P_0$ and $|s_{m,+}|^2 \approx 4.75
P_0$. However, if we instead keep $\tau_0 = \tau_p$ and increase
$\tau_m$ such that $\tau_m/\tau_0 = \tau_m/\tau_p = 10$, $U$ moves
only to $|s_{0,+}|^2 \approx 1.05 P_0$ and $|s_{m,+}|^2 \approx 0.27
P_0$. Note that, as mentioned previously, maximal stable conversion
efficiency is obtained for low signal power $|s_{m,+}|^2$ and input
power $|s_{0,+}|^2$ near the critical power $P_0$, regardless of
$\tau_k$. We note that rescaling $\beta$ simply scales the input power
and therefore changing $\beta$ does not affect the dynamics.

Thus far, we have focused on the up-conversion process: taking input
light at frequencies $\omega_0$ and $\omega_m$ and generating output
light at frequency $\omega_p > \omega_0$. However, it suffices to
consider the above system when $\Delta \omega < 0$ to understand the
physics of the alternative, down-conversion process: taking input
light at frequencies $\omega_0$ and $\omega_p$ and generating output
light at frequency $\omega_m$. For $\Delta \omega < 0$, we effectively
have $\omega_m \leftrightarrow \omega_p$. In this regime, all of the
above analysis holds, and in particular, the maximal efficiency, given
by \eqref{etamax}, is obtained as $|s_{m,+}|^2 \to 0$ with
$|s_{0,+}|^2 = P_0$. Similarly, the stability of the solutions follow
similar trends to those outlined above.

\subsection{$\Delta \omega \geq \omega_0$ regime: Complete conversion}
\label{sec:cc}

When $\Delta \omega$ is larger than $\omega_0$, we argued in
\secref{qlc-cc} that the system is capable of complete conversion,
i.e. $\eta = 1$. In this section, we demonstrate the existence of a
critical steady-state solution to the classical coupled mode equations
with complete conversion and analyze the stability of this critical
solution, as well as relate DFWM to our previous work on THG
\cite{Rodriguez07:OE,Hashemi09}.

\begin{figure}[!t]
\includegraphics[width=\columnwidth]{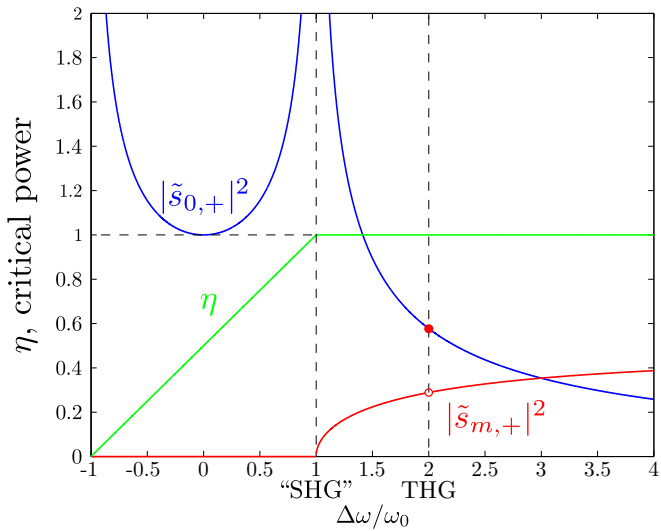}
\caption{(Color online) Plot of the critical powers
  $|\tilde{s}_{0,+}|^2$ (blue), $|\tilde{s}_{m,+}|^2$ (red), and
  maximum steady-state efficiency $\eta$ (green) as a function of
  $\Delta \omega/\omega_0$ (the tilde over the critical powers
  indicates that the values have been rescaled by the factor $4/\tau_0
  |\beta_0| \sqrt{\tau_m \tau_p \omega_0^2}$). The vertical dashed
  lines at $\Delta \omega = \omega_0$ and $\Delta \omega = 2 \omega_0$
  indicate special degenerate regimes, corresponding to ``second
  harmonic generation'' (SHG) and third harmonic generation
  (THG). (Note the discontinuity in $|\tilde{s}_{m,+}|^2$ located at
  $\Delta \omega = 2 \omega_0$, explained in the text).}
\label{fig:fig6}
\end{figure}

As in the previous section, we consider the equations of motion
\eqrefrange{cme1}{cme3} in the steady state. To obtain the critical
solution, we again require depletion of the pump power,
i.e. $s_{0,-}=0$. However, as argued in \secref{qlc-cc}, complete
depletion of the signal, $s_{m,-} = 0$ must also occur. Recall from
\secref{qlc-cc} that complete $\omega_m$ depletion is possible in the
$\Delta \omega \geq \omega_0$ regime since the up-conversion process
does not produce $\omega_m$ photons (see \figref{fig2}). Imposing the
depletion constraints on the steady-state equations of motion yields
the following critical cavity energies $|a_k^\crit|^2$:
\begin{align}
  |a_0^\crit|^2 &= \frac{1}{|\beta_m|\sqrt{\tau_m \tau_p |\omega_m \omega_p|}},  \label{eq:a0crit}\\
  |a_m^\crit|^2 &= \frac{\tau_m |\omega_m|}{2\tau_0 \omega_0} |a_0^\crit|^2, \label{eq:amcrit} \\
  |a_p^\crit|^2 &= \frac{\tau_p \omega_p}{2 \tau_0 \omega_0}
  |a_0^\crit|^2, \label{eq:apcrit}
\end{align}
which lead to the following critical powers:
\begin{align}
  |s_{0,+}^\crit|^2 &= P_0 \label{eq:s0crit}\\
  |s_{m,+}^\crit|^2 &= \frac{|\omega_m|}{2\omega_0}P_0, \label{eq:smcrit}
\end{align}
where $P_0$ is given by \eqref{p0}. Solving for the corresponding
output signal $|s_{p,-}|^2$, the output power is indeed 100\% of the
input power, as required by energy conservation. (In contrast, the
assumption that $s_{0,-} = s_{m,-} = 0$ in the $|\Delta \omega | <
\omega_0$ case yields no solution). Note that the critical signal power
$|s_{m,+}^\crit|^2$ is now non-zero, due to the fact that the energy
from the signal $\omega_m$ photons is necessary to produce the output
$\omega_p$ photons. This is in contrast with the $|\Delta \omega| <
\omega_0$ regime where maximal conversion efficiency was only achieved
in the limit as input signal power $|s_{m,+}|^2$ decreased to
zero. The critical pump and signal powers, with the corresponding
maximum efficiency $\eta$, are plotted versus $\Delta \omega$ in
\figref{fig6} for both $\Delta \omega$ regimes.

As may be noted from \figref{fig6}, there are two particular values of
$\Delta \omega$ that warrant special attention when $\Delta \omega
\geq \omega_0$. The first case, when $\Delta \omega = \omega_0$, the
``SHG'' case, was discussed in the previous section. The second case
is when $\Delta \omega = 2\omega_0$. In this case, $\omega_m = -
\omega_0$ and $\omega_p = 3 \omega_0$, reminiscent of third-harmonic
generation (THG). In fact, this case of DFWM corresponds exactly to
$\chithree$ THG, and thus $\Delta \omega > \omega_0$ strictly
generalizes our previous THG analysis \cite{Hashemi09}. To see this,
some care must be taken to adjust the coupling coefficients $\beta_k$
given in \eqrefrange{beta0}{betap} to properly implement the rotating
wave approximation; since $\omega_m = - \omega_0$, we have that $a_m =
a_0^*$, and thus $\beta_0 \to \beta_0 + \beta_m^*$ and $\beta_m \to
\beta_m + \beta_0^*$. This results in $\beta_0 = \beta_m^* = 3
\beta_p^*$, exactly as shown in \cite{Hashemi09}. Furthermore, we have
$|s_{0,+}^\crit|^2=|s_{m,+}^\crit|^2 = P_0$ (note that this differs by
a factor of two from \eqref{smcrit}, due to the adjusted $\beta_k$
values); upon requiring that $\tau_0 = \tau_m$, this recovers the
critical power previously obtained for THG \cite{Hashemi09}. Note that
the correspondence between $\Delta \omega = 2\omega_0$ and $\chithree$
THG is exact, whereas the $\Delta \omega = \omega_0$ limit has little
in common with $\chitwo$ SHG as discussed above.

\begin{figure}[t!]
\includegraphics[width=\columnwidth]{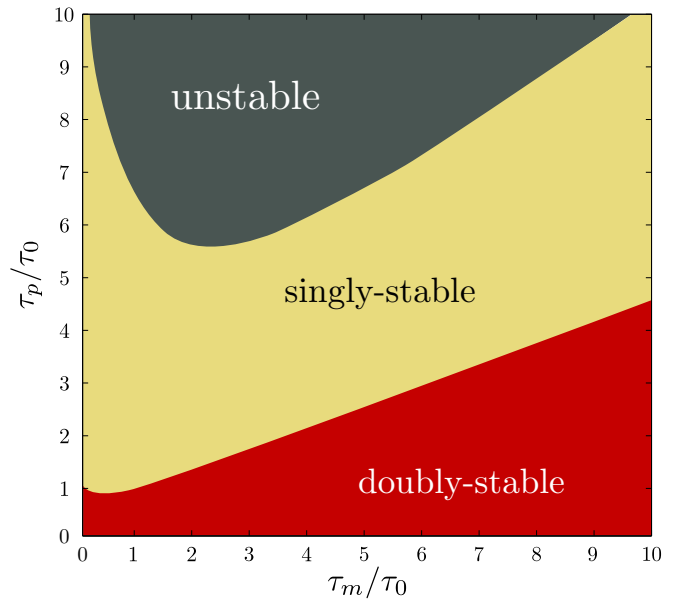}
\caption{(Color online) Stability contours (number of stable
  solutions) as a function of modal lifetimes $\tau_m$ and $\tau_p$,
  normalized by $\tau_0$, pumping at the critical input powers
  $|s_{0,+}^\crit|^2$ and $|s_{m,+}^\crit|^2$. The stability in the
  $\Delta \omega \geq \omega_0$ regime is independent of the value of
  $\Delta \omega$.}
\label{fig:fig7}
\end{figure}

The existence of an $s_{0,-}=s_{m,-} = 0$ solution having demonstrated
the existence of critical powers where 100\% conversion can be
achieved, we are now interested in characterizing the system at this
critical power by studying all of the fixed points. These fixed points
were obtained using Mathematica as in the previous section, and their
stability was determined via linear stability analysis as before. For
the critical input power, the steady-state equations of motion yield
three solutions; however, in contrast to the $|\Delta \omega| <
\omega_0$ regime, there exists multistability when $\Delta \omega \geq
\omega_0$. Similar to the case of THG ($\Delta \omega = \omega_0$),
the system is either singly stable, doubly stable, or unstable,
depending on the values of the mode lifetimes $\tau_k$ (see
\figref{fig7}). In this $\Delta \omega > \omega_0$ regime, the
stability of the solutions does not depend on $\Delta \omega$, again
in contrast with the quantum-limited regime. Unlike the $|\Delta
\omega| < \omega_0$ regime, the value of $\tau_m$ now plays a
significant role in the stability of the solutions.

\subsection{Self- and Cross-Phase Modulation $(\alpha \neq 0)$}
\label{sec:alpha}

Finally, we briefly consider the effects of SPM and XPM. This
corresponds to taking the coefficients $\alpha_{jk}$ to be non-zero;
as mentioned above, for simplicity, we take all the coefficients to be
equal, i.e. $\alpha_{jk} = \alpha$ for all $j,k$. The main effect of
SPM and XPM in \eqrefrange{cme1}{cme2} is to shift the resonant
frequencies of the cavity in proportion to the energy of the modes in
the cavity. Generally, this drives the frequency input light off
resonance and therefore degrades the overall conversion efficiency
obtained in \secreftwo{qlc}{cc}, as shown in
\citeasnoun{Hashemi09}. However, in \citeasnoun{Hashemi09}, we showed
that one simple way to overcome this difficulty is to pre-shift the
cavity resonant frequencies so as to compensate for the SPM/XPM
effects when operating near the critical input power.  Unfortunately,
this will inevitably affect the stability analysis obtained in the
$\alpha=0$ case, and therefore a new analysis that includes SPM/XPM
effects must be performed. In this remainder of this section,
we only analyze the stability of the maximal-efficiency solutions
obtained in \secreftwo{qlc}{cc}, and in particular, we find that 100\%
photon-conversion efficiency can be obtained in this case as well.

The change in cavity frequency due to SPM/XPM can be accounted for by
a pre-shifting technique described in \citeasnoun{Hashemi09}. In
particular, the $\alpha$ terms in \eqrefrange{cme1}{cme3} act to shift
the cavity resonant frequencies from $\omega^{\textnormal{cav}}_k \to
\omega_k^{\textnormal{NL}}$, spoiling the frequency-conservation
relations necessary for efficient nonlinear frequency conversion as
well as detuning the resonances from the input light.  However, one
can simply design the cavity frequencies to be resonant at the shifted
frequencies, i.e. $\omega_k^{\textnormal{cav}} =
\omega_k^{\textnormal{NL}}$, for a given steady-state solution. For
the critical solutions corresponding to 100\% photon-conversion
efficiency, this implies that the new cavity frequencies will be given
by \cite{Hashemi09}:
\begin{align}
  \omega_0^\cav &= \frac{\omega_0}{1-\alpha(|a_0^\crit|^2 +
    |a_m^\crit|^2 +
    |a_p^\crit|^2)} \label{eq:shift-freq1}\\ \omega_m^\cav &=
  \frac{\omega_m}{1-\alpha(|a_0^\crit|^2 + |a_m^\crit|^2 +
    |a_p^\crit|^2)} \label{eq:shift-freq2}\\ \omega_p^\cav &=
  \frac{\omega_p}{1-\alpha(|a_0^\crit|^2 + |a_m^\crit|^2 +
    |a_p^\crit|^2)}.\label{eq:shift-freq3},
\end{align}
where $|a^{\crit}_k|^2$ are the energies of the modes at critical power. For
cavities resonances $\omega_k^\cav$, the new equations of motion are
given by:
\begin{align}
  \frac{da_0}{dt} = &\Big[ i \omega_0^\cav (1-\alpha_{00}|a_0|^2 - \alpha_{0m}
  |a_m|^2 - \alpha_{0p} |a_p|^2) \nonumber \\ &- \left. \frac{1}{\tau_0} \right] a_0 - i
  \omega_0 \beta_0 a_0^* a_m a_p +
  \sqrt{\frac{2}{\tau_{s,0}}}s_{0,+} \label{eq:cme21} \\
\frac{da_m}{dt} = &\Big[ i \omega_m^\cav (1-\alpha_{m0}|a_0|^2 - \alpha_{mm}
|a_m|^2 - \alpha_{mp} |a_p|^2) \nonumber \\ &- \left. \frac{1}{\tau_m} \right] a_m -
i\omega_m \beta_m a_0^2 a_p^* +
\sqrt{\frac{2}{\tau_{s,m}}}s_{m,+} \label{eq:cme22} \\
\frac{da_p}{dt} = &\Big[ i \omega_p^\cav (1-\alpha_{p0}|a_0|^2 - \alpha_{pm}
|a_m|^2 - \alpha_{pp} |a_p|^2) \nonumber \\ &- \left. \frac{1}{\tau_p} \right] a_p  -
i\omega_p \beta_p a_0^2 a_m^* \label{eq:cme23},
\end{align}
Note that the frequencies $\omega_k$ multiplying the $\beta_k$ terms
do not need to be shifted, since the terms introduced by such a
shifting will be higher order in $\chi^{(3)}$. By inspection, we
observe that the solutions obtained in \secreftwo{qlc}{cc} at critical
input power $a_k^\crit$ are also solutions of
\eqrefrange{cme21}{cme22}, but as explained above, their stability may
change.  Using the results from \secreftwo{qlc}{cc}, we now study the
stability properties of these solutions in the two $\Delta \omega$
regimes.

\begin{figure}[t]
  \centering
  \includegraphics[width=\columnwidth]{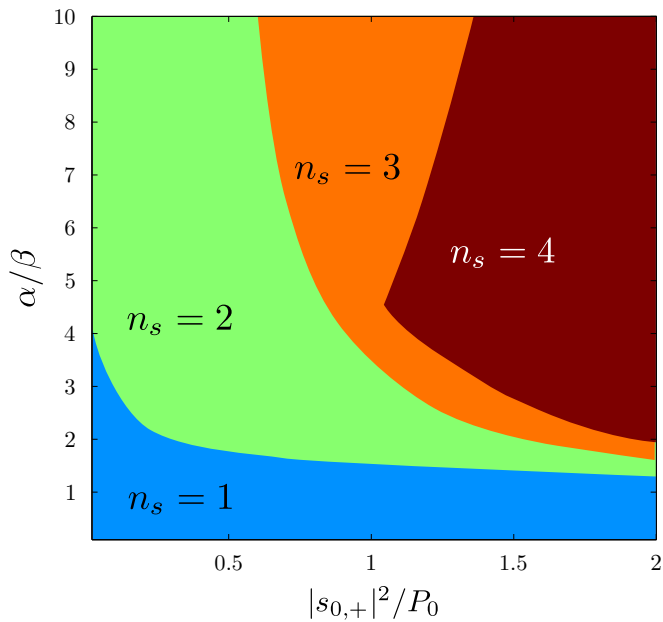}
  \caption{(Color online) Contour plot of number of stable solutions
    ($n_s$) as a function of $\alpha/\beta$ and $|s_{0,+}|^2/P_0$, for
    input pump power $|s_{m,+}|^2=0.1 P_0$, and for the system
    described in the text.}
  \label{fig:fig8}
\end{figure}

We first consider the $\Delta \omega \leq \omega_0$ regime. As in
\secref{qlc}, we restrict our analysis to a specific parameter regime
($\tau_0 = \tau_m = \tau_p = 100/\omega_0$, $\beta=10^{-4}$, and
$\Delta \omega = 0.05 \omega_0$) for simplicity, although our
qualitative conclusions apply to other parameter ranges. As discussed
above in \secref{qlc}, the maximal efficiency is obtained for input
light with $|s_{0,+}|^2 = P_0$ as $|s_{m,+}|^2\to 0$. Since one must
always pump with finite $|s_{m,+}|^2$, and there are no analytic
solutions in this case, we solve for the field energies
$|a_k^\crit|^2$ numerically at a small $|s_{m,+}|^2$ and for
$|s_{0,+}|^2 = P_0$ in the case of $\alpha=0$ in order to compute the
shifted frequencies \eqrefrange{shift-freq1}{shift-freq3}. This allows
us to solve the coupled-mode equations \eqrefrange{cme21}{cme22} and
therefore obtain the steady-state field amplitudes and phases. As in
\citeasnoun{Hashemi09}, the inclusion of self- and cross-phase
modulation introduces new steady-state solutions absent in the
$\alpha=0$ case, and the stability of the old and new solutions are
then examined again via a linear stability analysis, as in
\secref{qlc}. In particular, we find that the inclusion of SPM/XPM
does not destroy the stability of the maximal efficiency solution in
the $\alpha=0$ case studied in \secref{qlc}, and in fact creates
additional stable solutions, as shown in \figref{fig8}.

A similar analysis can be performed in the $\Delta \omega>\omega_0$
regime, where it is possible to obtain the analytic form of the
maximal efficiency solutions [\eqrefrange{a0crit}{apcrit} in
  \secref{cc}]. We find that, as in the previous regime, the presence
of $\alpha$ introduces additional stable solutions, while retaining
the original 100\% efficiency $\alpha=0$ solution, over finite regions
of the parameter space.

The presence of SPM/XPM in our system provides an opportunity to
observe rich and interesting dynamical behaviors, including limit
cycles and hysteresis effects, that we do not explore in this
paper. As noted in this section, the inclusion of these effects is not
prohibitive for 100\% nonlinear frequency conversion although
predicting which parameter regimes allow for such conversion will
depend on the system under question. In the future, we plan to examine
SPM/XPM effects in more detail for realistic geometries with realistic
values of $\alpha_{ij}$ and $\beta_i$. As in \citeasnoun{Hashemi09},
the presence of multiple stable solutions means that the manner in
which the source is initiated will determine which solution is
excited, but a simple initialization procedure similar to that in
\citeasnoun{Hashemi09} should be possible to excite the
maximal-efficiency solution.

\section{Conclusion}

By exploiting a simple but rigorous coupled-mode theory framework, we
have demonstrated the possibility of achieving highly-efficient
(low-power) DWFM in triply-resonant cavities, similar to our previous
work in SHG and THG \cite{Rodriguez07:OE,Hashemi09}. We conclude that
there are two main regimes of operation, determined by the ratio of
$\Delta \omega$ to $\omega_0$. In particular, whereas the maximal
efficiency obtainable in the $\Delta \omega \leq \omega_0$ regime,
corresponding to conversion between closely-spaced resonances, is
bounded above by a quantum-limited process, there is no such bound
when $\Delta \omega > \omega_0$. In both regimes, a suitable choice of
system parameters leads to stable, maximal-efficiency nonlinear
frequency conversion, even in the presence of SPM and XPM effects. We
remark that all of the results obtained in this paper correspond to
the idealized case of lossless interactions, since the main focus of
the paper is in examining the basic considerations involved in
operating with these systems rather than predicting results for
specific experimentally-relevant systems. Nevertheless, based on our
previous experience with SHG and THG \cite{Rodriguez07:OE,Hashemi09},
we expect that linear and nonlinear losses, e.g. coming from radiation
or material absorption, will only act to slightly decrease the overall
conversion efficiency and will not affect the qualitative predictions
here. In a future manuscript, we plan to explore DFWM in a realistic
geometry such as a ring resonator coupled to an index-guided waveguide
and study some of the dynamical effects arising from SPM/XPM.

\subsection*{Acknowledgments}
This work was supported by the MRSEC Program of the NSF under Award
No. DMR-0819762 and by the U.S.A.R.O. through the ISN under Contract
No. W911NF-07-D-0004. We are also grateful to Jorge Bravo-Abad at
U.~Autonoma de Madrid for helpful discussions.

\bibliography{photon}
\end{document}